# Exact Performance Analysis of Partial Relay Selection Based on Shadowing Side Information over Generalized Composite Fading Channels

Ferkan Yilmaz, *Member, IEEE* and Mohamed-Slim Alouini, *Fellow, IEEE*

## Abstract

Channel side information (CSI) information varies very fast in the millimeter wave radio frequencies so its coherent time is much smaller than that of the CSI in low-frequencies (about 6 GHz or below) under the same mobility conditions. Further, when the feedback channels, regardless of being wireless or wireline, have a backhaul / transmission delay which is practically so much higher than the coherent time of the CSI in the millimeter wave radio frequencies. As such in practical scenarios, the CSI will be inherently changed in the fading environment before successfully reaches to the central entity (CE) through feedback channels, which specifically means that the CSI reached at the CE certainly becomes inaccurate, and therefore the CSI cannot be promptly and successfully used for relay selection protocols (RSPs). Correspondingly, the system performance drastically deteriorates. However, the shadowing side information (SSI) varies more slowly such that its coherent time is substantially greater than both that of the CSI and backhaul / transmission delays in feedback channels. Within this concept to circumvent these undesirable limitations, we suggest in this paper that some performance gain can be still obtained utilizing only the SSI for RSPs. In contrast to the other RSPs available in the literature, we propose a partial-RSP in dual-hop amplify-and-forward relaying system utilizing only the SSI of the relay terminals instead of their CSI. Furthermore, we offer a unified and generic performance expression of the proposed partial-RSP, spontaneously combining the analyzes of the average bit error probability, ergodic capacity, and moment generating function as a single analysis. Finally, as an illustration of the mathematical formalism, some numerical and simulation results are carried out for the generalized composite fading environments whose fading envelope follows extended generalized-K distribution, and these numerical and simulation results are shown to be in perfect agreement.

Ferkan Yilmaz and Mohamed-Slim Alouini are with King Abdullah University of Science and Technology (KAUST), Al-Khawarizmi Applied Math. Building, Thuwal 23955-6900, Makkah Province, Kingdom of Saudi Arabia (e-mail: {ferkan.yilmaz, slim.alouini}@kaust.edu.sa).

This work was supported by King Abdullah University of Science and Technology (KAUST).

# I. INTRODUCTION

RELAY technology has recently gained great interest amongst researchers, engineers and developers in both academic and industrial fields as a transmission technique promising in millimeter wave (60 GHz or above) radio frequencies, improving the quality of service, providing high data rate, and extending the coverage area without additional transmit power in deeply shadowed (composite) fading environments [1]–[3]. More specifically, referring to the current research and development projections we are currently dealing with, it is reasonably estimated that in future wireless technologies (5G and the beyond), there will be explicitly many wireless high-speed computers / devices / phones using external central process unit (CPU), external memory and external storage trough the medium of wireless communications links while their sizes are getting smaller and smaller. These applications naturally require high wireless data transfer with sophisticated usage of millimeter wave radio frequencies. Due to severe propagation characteristics in millimeter wave radio frequencies, the coverage area is generally within a $10\,\mathrm{m}$ or much smaller radius. Accordingly, wireless communications systems require to be assisted by relay technology such that the transmission between the source terminal (ST) and the destination terminal (DT) is successfully achieved by one or more relay terminals (RTs). There exist two famous relaying techniques one of which is the amplify-and-forward (AF) technique [3] in which each RT simply amplifies the received signal and then forwards the amplified signal to the DT. The other one is the decode-and-forward (DF) technique [4] in which each RT decodes the received signal then forwards the decoded signal to the DT. While most frequently using these two relaying techniques, the purpose of relay technology is not only to increase the quality in signal transmission by obtaining spatial diversity at the DT via mutually independent copy of the transmitted information [1]–[3], [5] but also to increase the reliability of the signal transmission by reducing the fluctuations in the signalling throughput [6]. For that reason, more than one RT can be employed in the signal transmission [7]–[10], but this solution considerably increases the network complexity with respect to the channel estimation, signal detection and transmission synchronization as well as information combining, resource allocation, scheduling and power control for signal transmission [11]. In addition, as the number of RTs increases, the signal processing load at the central entity (CE) and DT drastically increases [11]. In order to maintain the corresponding network complexity as simple as possible while increasing the overall quality of service and keeping the acceptable reliability, the best known solution is to select only one RT which is in the best fading conditions to help the signal transmission from ST to the DT [3], [11]–[13].

Designing a wireless communications system based on a relay technology considerably depends on which relay selection protocol (RSP) is utilized. Specifically, the various types of RSPs are well studied and discussed in the literature for different fading environments [7]–[11], [13]–[21]. These protocols can be grouped as best-relay selection [7]–[9], [13], [14], [18], best-worse-channel selection [7], best-harmonic-mean selection [7], [14], nearest-neighbor selection [11], [15], reactive-relay selection [19], and partial relay selection [16]. In more details, the best-relay selection suggests that the RT that can achieve the highest signal-to-noise ratio (SNR) at the DT is chosen from all the available RTs. For instance, [14] studies the error probability performance of this protocol, and an upper bound is presented in a closed-form for Rayleigh fading environments. The other RSP is the best-worse-channel selection [10], [17] in which the RT with the largest of the worst channels from either ST $\to$ RT or RT $\to$ DT transmission links is selected among all the available RTs. The outage performance analysis for this protocol is successfully carried in both [17] and [10] for Rayleigh fading channels. It is further worth mentioning that both best-relay selection and best-worse-channel selection protocols achieve full diversity order. The other RSP achieving full diversity order is the best-harmonic-mean selection [7], [14] in which the RT that has the largest harmonic-mean of the SNRs of the ST $\to$ RT and RT $\to$ DT fading channels is selected. It is important to note that in order to implement these RSPs, there needs to be a CE collecting all channel state information (CSI) of all available transmission channels from ST $\to$ RTs and RTs $\to$ DT. Specifically, as the number of the RTs increases, the CE is drastically overloaded with all CSI information and then these RSPs, which maintain full-diversity order, become unfortunately impractical.

In order to simplify the selection complexity and reduce the overload at the CE, the nearest-neighbor selection is proposed in [11], [15] in which the RT nearest to the DT is selected, where "the nearest RT" is not necessarily the spatially nearest RT to the ST or DT but the RT with the strongest transmission channel (i.e., with the transmission channel having less amount-of-fading and less shadowing/path-loss) to the RT or DT. In addition to the nearest-neighbor selection, the other complexity-reduced RSPs such as reactive-RSP and partial-RSP were proposed in [16], [19]–[21], respectively. Note that both reactive-RSP and partial-RSP require the CSI information of only one of channels ST $\to$ RT and RT $\to$ DT, respectively. Hence, the complexity of the network is substantially reduced by these two protocols. More specifically, the reactive selection is an quite efficient solution especially when RTs are close to the ST, and in which the RT is selected achieving the highest instantaneous SNR at the RT $\to$ DT channel [19]. On the other hand, the partial-RSP [16], [20], [21] in which the RT with the highest SNR at the ST $\to$ RT channel is selected turns into an efficient solution when RTs are close to the DT.

As mentioned above, there exists a need in all RSPs to send all required CSI information to a CE through feedback channels in order to successfully achieve relay selection. However in practical cases in millimeter wave radio frequencies, there are some certain limitations for sending all required CSI information. For example,

i) The number of available feedback channels in a communication system is typically limited and generally smaller than the number of RTs, which inherently turns into a feedback channel assignment problem specially when the coherent time of the CSI information is substantially small,

ii) As the coherent time of the CSI information gets smaller, the energy consumption in the feedback channels significantly grows due to the throughput quality requirements.

iii) The feedback channels, regardless of being wireless or wireline, have a backhaul/transmission delay which is practically so much higher than the coherent time of the CSI in the millimeter wave radio frequencies. As such in practical applications, the CSI will be inherently altered in the fading channel before successfully reaches to the CE through feedback channels. Specifically meaning that under these specified conditions, the CSI in the fading channel and the CSI reached at the CE will be assuredly uncorrelated. Thus, the CSI received by the CE through feedback channels explicitly becomes inaccurate for the RSPs, and consequently the system performance drastically and distinctively deteriorates.

It is particularly useful to emphasize that these limitations even get worse as the mobility of the ST, RTs and DT increases. For example, increasing the carrier frequency in wireless systems proportionally and apparently magnifies the Doppler effects [22]. Following that the Doppler frequency in $60$ GHz is typically at least $10$ times greater than that in $6$ GHz frequencies under the same mobility conditions [22]–[24], which clearly states that there exists a proportional relation between Doppler's effect and the time-varying nature of the fading environments. Thus, the rate of the CSI information in $60$ GHz frequencies is certainly at least 10 times greater than that in $6$ GHz frequencies [25]. In other words, the CSI varies very fast so its coherent time is much smaller than that of a CSI in low-frequencies (about $6$ GHz or below) under the same mobility conditions. It is therefore not difficult to envision such a practical scenario in millimeter wave radio frequencies that all RTs will continuously try to use at least one feedback channels but some RTs could not be assigned to feedback channels because of the limited number of the feedback channels. In such a case, the RTs have to reduce their CSI feedback to a low rate in order to share the feedback channels amongst themselves. Because of reducing the feedback of the CSI, the CE will not correctly achieve the RSP, and the overall performance will take a turn for the worse. In particular under the circumstances when the CSIs of the RTs are non-identically distributed, the performance of the employed RSP is bounded by that of the average power (AP)-based RSP (i.e., pathloss-based RSP) in which the RT having the highest AP/the lowest pathloss is chosen from all RTS [11]. Otherwise when CSIs are identically distributed, it is bounded by that of the round-robin (RR)-based RSP in which each RT is periodically chosen for signal transmission.

In addition to the feedback channel assignment problem, the feedback channels, regardless of being wireless or wireline, have a backhaul/transmission delay which is practically so much higher than the coherent time of the CSI in the millimeter wave radio frequencies. As such in practical scenarios, the CSI will be inherently changed in the fading environment before successfully reaches to the central entity (CE) through feedback channels. Under these circumstances, even if there are enough number of feedback channels, the CSI in the fading channel and the CSI reached at the CE will not be same, and even within this practical scenario, they will certainly not be correlated. Consequently, the CSI cannot be promptly and successfully used for RSPs and the system performance explicitly and drastically deteriorates. Within this context, it is worthy noting that the shadowing side information (SSI) of a fading channel varies very slowly when compared with the CSI such that its coherent time is substantially greater than that of the CSI and backhaul/transmission delays in feedback channels. Specifically meaning that even if the CSI is surely difficult to be utilized for a RSP in millimeter wave radio frequencies, the SSI of the fading channel could be at least easily used to obtain some performance gain. As such, the relay selection based on the SSI information appropriately turns into an interesting solution for wireless communications systems operating in wireless millimeter wave frequencies. In particular, in contrast to the available RSPs [7]–[11], [13]–[21], we propose in this paper a new RSP based on the SSI instead of CSI, and we analyze its performance measures by offering a unified and generic performance expression that shows how to treat different performance analyzes as a single performance analysis. Explicitly, the followed analysis spontaneously combines the average bit error probability (ABEP), ergodic capacity (EC), and moment generating function (MGF) for a variety of fading environments. In this context, we reintroduce the extended generalized-K (EGK) distribution [26], [27] as a unified model for shadowed fading channels and then give some computationally efficient analytical results for performance measures.

The remainder of the paper is organized as follows. In Section II, the system model of the partial-

RSP in dual-hop AF relaying system that only employs the SSI is overall described in details for generalized composite fading environments,Then, the unified and generic performance analysis is presented in Section III such that some main results regarding the special cases of the proposed unified and generic performance expression are briefly discussed. For possible applications in composite fading environments, the EGK distribution is reintroduced in Section IV as a generalized composite distribution that provides a unified theory to model the envelope statistics of wireless communication channels commonly used in the literature. In the following, its cumulative distribution function (CDF) and reciprocal moment generating function (MGF) are obtained in a closed form. Then for EGK fading environments, the correctness and accuracy of the unified performance analysis have been numerically checked. Finally, the conclusions are drawn in the last section.

## II. PARTIAL RELAY SELECTION BASED ON SHADOWING SIDE INFORMATION

We consider an AF dual-hop transmission in millimeter wave radio frequencies, between one ST $S$ and one DT $D$ through the $L \geq 1$ number of RTs $\{R_1, R_2, \ldots, R_L\}$ as seen in Fig. 1. Assumed that there does not exist a direct link between the ST $S$ and DT $D$ due to the unsatisfactory channel quality, then the transmission can be performed only via RTs. In addition, appropriately assumed that the $L$ RTs are relatively close to the destination. However as mentioned before, due to the impediment of the limited number of feedback channels, the backhaul/transmission delays subsisting in feedback channels, and the difficulties in synchronization and power control among RTs, only one RT having the best conditions is selected to enable the transmission between the ST and DT. More precisely as seen in Fig. 1, for all hops $k \in \{1, 2\}$ and RTs $\ell \in \{1, 2, \ldots, L\}$, $\{\gamma_{k,\ell}\}$ denote the composite fading coefficients defined as the product of shadowing $\{\mathcal{S}_{k,\ell}\}$ and multipath fast fading $\{\mathcal{G}_{k,\ell}\}$ such that $\gamma_{k,\ell} = \mathcal{S}_{k,\ell} \mathcal{G}_{k,\ell}$. Without loss of generality, the power of multipath fast fading, i.e., $\mathbb{E}[\mathcal{G}_{k,\ell}]$ is assumed to be unity, where $\mathbb{E}[\cdot]$ denotes the expectation operator. Hence, the short-term and long-term average powers of the composite fading $\gamma_{k,\ell}$ are $\mathbb{E}[\gamma_{k,\ell}] = \mathcal{S}_{k,\ell}$ and $\mathbb{E}[\gamma_{k,\ell}] = \mathbb{E}[\mathcal{S}_{k,\ell}] = \Omega_{k,\ell}$, respectively. Accordingly within this practical scenario, a RT could be in general picked from a set of RTs for the best transmission, archetypically by means of utilizing either CSI composed of shadowing and multipath fast fading components.

However previously mentioned some certain limitations are, we propose a partial-RSP based upon the SSI comprising only shadowing component for communications systems operating in millimeter wave radio frequencies. Explicitly within this proposed partial-RSP, the ST $S$ broadcasts predetermined training symbols to all RTs $\{R_1, R_2, \ldots, R_L\}$. Then, for all $\ell \in \{1, 2, \ldots, L\}$, the $\ell$th RT estimates the SSI $\mathcal{S}_{1,\ell}$ with the aid of training symbols recovered from the transmission channel from the ST $S$. As a sequel using the feedback channels, each RT sends its SSI information to the CE which selects the RT having the highest SSI as

$$M = \arg \max_{\ell \in \{1,2,\ldots,L\}} \mathcal{S}_{1,\ell} \tag{1}$$

where $M \in \{1, 2, \ldots, L\}$ is the index of the selected RT. It is important to note that the index of the selected RT, i.e., $M$ is a discrete random variable (RV) because of random nature of the SSI variation, and its probability mass function (PMF) can be simply written as

$$p_M(m) = \sum_{\ell=1}^{L} \mu_\ell \, \delta(m - \ell), \tag{2}$$

where $\delta(\cdot)$ is the Dirac's delta function [28, Eq.(1.8.1)], and where the coefficient $\mu_\ell$ denotes the probability of that the $\ell$th RT is selected, and it is namely defined as $\mu_\ell \triangleq \Pr[M = \ell]$, i.e.,

$$\mu_\ell = \int_0^\infty p_{\mathcal{S}_{1,\ell}}(s) \prod_{\substack{k=1 \\ k \neq \ell}}^{L} P_{\mathcal{S}_{1,k}}(s) \, ds, \tag{3}$$

such that $\sum_{\ell=1}^{L} \mu_\ell = 1$, where for $s \in \mathbb{R}^+$, both $p_{\mathcal{S}_{1,k}}(s)$ and $P_{\mathcal{S}_{1,k}}(s)$ respectively denote the probability density function (PDF) and CDF of the SSI $\mathcal{S}_{1,k}$ that the first-hop transmission is subjected to. Under the condition of that the $M$th RT is selected, the ST $S$ transmits the signal with a unit average power and the $M$th RT receives the signal and amplifies it with the gain

$$\mathcal{G}_M = \frac{1}{\sqrt{\widehat{\gamma}_{1,M}}} = \frac{1}{\sqrt{\widehat{\mathcal{S}}_{1,M} \, \mathcal{G}_{1,M}}}, \tag{4}$$

where $\widehat{\mathcal{S}}_{1,M} \triangleq \max_{k \in \{1,2,\ldots,L\}} \mathcal{S}_{1,k}$ denotes the shadowing of the first-hop fading channel conditioned on that the $M$th RT is selected. In this sense, the instantaneous SNR of the first-hop transmission becomes $\widehat{\gamma}_{1,M} = \widehat{\mathcal{S}}_{1,M} \, \mathcal{G}_{1,M}$. Then, remembering that the index of the selected RT $M$ is a discrete RV with the PMF given in (2), the overall instantaneous SNR $\gamma_{end}$ at the DT $D$ can be written in terms of a hyper-distribution, that is

$$\gamma_{end} = \sum_{\ell=1}^{L} \frac{\widehat{\gamma}_{1,\ell} \, \gamma_{2,\ell}}{\widehat{\gamma}_{1,\ell} + \gamma_{2,\ell}} \delta_{M,\ell}, \tag{5}$$

where $\widehat{\gamma}_{1,\ell} = \widehat{\mathcal{S}}_{1,\ell} \, \mathcal{G}_{1,\ell}$ denotes the instantaneous SNR of the first-hop transmission conditioned on that the $\ell$th RT is selected, and where $\delta_{n,k}$ is the Kronecker's delta [28, Eq.(5.10.1.3)] such that $\delta_{n,k} = 1$ if $n = k$ and $\delta_{n,k} = 0$ otherwise. As mentioned before, the RTs are quite close to the DT $D$, creating a relay cluster. In this case, it is appropriate to assume that the transmission channels from RTs $R_1, R_2, \ldots, R_L$ to the DT $D$ are subject to negligible shadowing effect, which means that for all $\ell \in \{1, 2, \ldots, L\}$, the variance of the shadowing becomes $\text{Var}\left[\mathcal{S}_{2,\ell}\right] \approx 0$, where $\text{Var}\left[\cdot\right]$ is the variance operator, and its PDF $p_{\mathcal{S}_{2,\ell}}(s)$ turns into a Dirac's delta distribution $p_{\mathcal{S}_{2,\ell}}(s) = \delta(s - \Omega_{2,\ell})$. With this inference, the overall instantaneous SNR $\gamma_{end}$ can be easily given as

$$\gamma_{end} = \sum_{\ell=1}^{L} \frac{\widehat{\gamma}_{1,\ell}\,\Omega_{2,\ell}\mathcal{G}_{2,\ell}}{\widehat{\gamma}_{1,\ell} + \Omega_{2,\ell}\mathcal{G}_{2,\ell}} \delta_{M,\ell}. \tag{6}$$

In the next sections, the performance of the SSI-based partial-RSP is analyzed, and some key results are presented with the discussion for generalized composite fading channels.

### III. EXACT PERFORMANCE ANALYSIS OVER GENERALIZED FADING CHANNELS

In this section, we study the exact performance analysis of the partial-RSP, depicted in Fig. 1, using the SSI of all fading channels from the ST $S$ to all RTs $R_1, R_2, \ldots, R_L$.

#### A. Unified Performance Analysis

The ABEP and EC are two important performance measures to quantify the achievable performance gain with the aid of the SSI-based partial-RSP. Over the past four decades, these ABEP and EC performance measures have been considered as two different problems. As such for these two performance measures, many separate and different solutions have been proposed in the literature for a variety of modulation schemes, diversity combining techniques in fading environments. The authors have recently offered in [29] a generic unified performance (UP) expression that mathematically combines the ABEP and EC performance measures in a tractable way for a certain value of instantaneous SNR $\gamma_{end}$, that is

$$\mathcal{P}_{UP}(\gamma_{end}) = c - \frac{n}{2}\left\{c - \frac{(-1)^n c}{\Gamma(b)}\,\text{G}_{n,2}^{1,n}\!\left[a\gamma_{end}\,\middle|\,\begin{matrix}\Lambda_1^{(n)}\\b,0\end{matrix}\right]\right\}, \tag{7}$$

where $\Gamma(\cdot)$ is the Gamma function [30, Eq. (6.5.3)] and $\text{G}_{p,q}^{m,n}[\cdot]$ denotes the Meijer's G function [31, Eq. (8.3.22)], and where the coefficient set $\Lambda_a^{(n)}$, $n \in \mathbb{N}$ is defined as

$$\Lambda_a^{(n)} \equiv \overbrace{a, a, \ldots, a}^{n\text{ times}}. \tag{8}$$

It is informative to note that substituting $n = 1$ and $c = 1$ into (7) results in the well-known conditional bit error probability (BEP) $\mathcal{P}_{BEP}(\gamma_{end})$ proposed by Wojnar in [32, Eq. (13)], that is

$$\mathcal{P}_{BEP}(\gamma_{end}) = \frac{\Gamma(b, a\gamma_{end})}{2\Gamma(b)}, \quad a, b \in \left\{1, \frac{1}{2}\right\}, \tag{9}$$

where $\Gamma(\cdot, \cdot)$ denotes the lower incomplete Gamma function [33, Eqs. (8.350/2)], and where $a$ depends on the type of modulation ($\frac{1}{2}$ for orthogonal frequency shift keying (FSK), 1 for antipodal phase shift keying (PSK)), $b$ depends on the type of detection ($\frac{1}{2}$ for coherent, 1 for non-coherent). Additionally noting that for a fading channel with the bandwidth $W$ Hz, substituting $a=1$, $b=1$, $c=W/\log(2)$ and $n=2$ into (7) where $\log(\cdot)$ is the natural logarithm, (7) simplifies into the conditional channel capacity $\mathcal{P}_C(\gamma_{end})$ in bits/s/Hz, that is

$$\mathcal{P}_C(\gamma_{end}) = W\log_2(1+\gamma_{end}), \quad \text{bits/s/Hz}, \tag{10}$$

where $\log_2(\cdot)$ is the binary logarithm.

It is additionally useful to mention that the instantaneous BEP $\mathcal{P}_{BEP}(\gamma_{end})$ given by (9) simplifies to $\mathcal{P}_{BEP}(\gamma_{end}) = \frac{1}{2}\exp(-a\gamma_{end})$ for non-coherent binary modulation schemes. With this inference, the unified expression $\mathcal{P}_{UP}(\gamma_{end})$ given by (7) can also be employed to obtain the MGF of the instantaneous SNR $\gamma_{end}$. More specifically, setting $a=p$ with $p \in \mathbb{C} \wedge \Re\{p\} \in \mathbb{R}^+$, $b=1$, $c=2$ and $n=1$ in (7) simplifies into the instantaneous MGF, i.e.,

$$\mathcal{P}_{MGF}(\gamma_{end}) = \exp(-p\gamma_{end}), \tag{11}$$

which enables the utilization of (7) so as to obtain the statistical characterizations such as PDF, CDF and moments within the same performance analysis framework.

Eventually, using the unified performance expression $\mathcal{P}_{UP}(\gamma_{end})$ given in (7), the average unified performance (AUP) $\mathcal{P}_{AUP} = \mathbb{E}[\mathcal{P}_{UP}(\gamma_{end})]$ of the partial-RSP based on the SSI of the first-hop fading channels available between the ST $S$ and the RTs $R_1, R_2, \ldots, R_L$ is given in the following theorem.

**Theorem 1** (Unified Performance). *The AUP $\mathcal{P}_{AUP}$ of the SSI-based partial-RSP is given by*

$$\mathcal{P}_{AUP} = -\int_0^\infty \mathcal{Z}_{a,b,c}^n(u) \sum_{\ell=1}^L \mu_\ell \left\{ \mathcal{M}_{1/\widehat{\gamma}_{1,\ell}}(u) \frac{\partial}{\partial u}\mathcal{M}_{1/\gamma_{2,\ell}}(u) + \mathcal{M}_{1/\gamma_{2,\ell}}(u) \frac{\partial}{\partial u}\mathcal{M}_{1/\widehat{\gamma}_{1,\ell}}(u) \right\} du, \tag{12}$$

*where the auxiliary function $\mathcal{Z}_{a,b}^n(u)$ is defined as*

$$\mathcal{Z}_{a,b,c}^n(u) = c - \frac{n}{2}\left\{ c - \frac{(-1)^n c}{\Gamma(b)} G_{n,3}^{1,n}\left[ au \,\bigg|\, \begin{matrix} \Lambda_1^{(n)} \\ b, 0, 0 \end{matrix} \right] \right\}. \tag{13}$$

*where $a$, $b$, $c$ and $n$ are the performance parameters explained before. Furthermore in (12), for all $\ell \in \{1, 2, \ldots, L\}$, the coefficient $\mu_\ell$ is given in (3) denoting the probability of that the $\ell$th RT is selected. Further, both $\mathcal{M}_{1/\widehat{\gamma}_{1,\ell}}(p) = \mathbb{E}[\exp(-p/\widehat{\gamma}_{1,\ell})]$ and $\mathcal{M}_{2/\gamma_{2,\ell}}(p) = \mathbb{E}[\exp(-p/\gamma_{2,\ell})]$ denote the reciprocal MGFs for the first and second hops' instantaneous SNRs $\widehat{\gamma}_{1,\ell}$ and $\gamma_{2,\ell}$.*

*Proof:* Explicitly for a certain non-negative distribution of the instantaneous SNR $\gamma_{end}$, the AUP $\mathcal{P}_{AUP} = \mathbb{E}[\mathcal{P}_{UP}(\gamma_{end})]$ is completely represented as

$$\mathcal{P}_{AUP} = c - \frac{n}{2}\left\{ c + \frac{(-1)^n c}{\Gamma(b)} \int_0^\infty G_{n+1,2}^{1,n}\left[ \frac{a}{p} \,\bigg|\, \begin{matrix} \Lambda_1^{(n)}, 1 \\ b, 0 \end{matrix} \right]\left[ \frac{\partial}{\partial p}\mathcal{M}_{\gamma_{end}}(p) \right] dp \right\}, \tag{14}$$

using the MGF-approach given in [29, Eq. (30)]. Furthermore in (14), $\mathcal{M}_{\gamma_{end}}(p)$, $p \in \mathbb{R}^+$ denotes the MGF of the overall instantaneous SNR $\gamma_{end}$, and is in general impossible to be in closed form because of the nature of composite fading channels. However in order to reduce (14) into a computationally efficient single-integral, $\mathcal{M}_{\gamma_{end}}(p)$ is required in a closed-form. Averaging with the PMF given by (2), the overall MGF $\mathcal{M}_{\gamma_{end}}(p) = \mathbb{E}\left[e^{-p\,\gamma_{end}}\right]$ can be written as

$$\mathcal{M}_{\gamma_{end}}(p) = \sum_{\ell=1}^{L} \Pr\left[M = \ell\right] \mathbb{E}\left[e^{-p\,\gamma_{end}}\,\middle|\, M = \ell\right], \quad (15a)$$

$$= \sum_{\ell=1}^{L} \mu_\ell \, \mathbb{E}\left[\exp\left(-p \frac{\widehat{\gamma}_{1,\ell}\,\gamma_{2,\ell}}{\widehat{\gamma}_{1,\ell} + \gamma_{2,\ell}}\right)\right], \quad (15b)$$

where for all $\ell \in \{1, 2, \ldots, L\}$, the coefficient $\mu_\ell$ which denotes the probability of that the $\ell$th RT is selected is given in (3). Note that the instantaneous SNR at the DT becomes the normalized harmonic mean of the first and second hops' instantaneous SNRs $\widehat{\gamma}_{1,\ell}$ and $\gamma_{2,\ell}$ when the $\ell$th RT is selected. More precisely, after performing some algebraic manipulations using [34, Theorem 3], $\mathcal{M}_{\gamma_{end}}(p)$ is obtained as

$$\mathcal{M}_{\gamma_{end}}(p) = -\int_0^\infty J_0\left(2\sqrt{pu}\right) \sum_{\ell=1}^{L} \mu_\ell \left\{\mathcal{M}_{1/\widehat{\gamma}_{1,\ell}}(u)\frac{\partial}{\partial u}\mathcal{M}_{1/\gamma_{2,\ell}}(u) + \mathcal{M}_{1/\gamma_{2,\ell}}(u)\frac{\partial}{\partial u}\mathcal{M}_{1/\widehat{\gamma}_{1,\ell}}(u)\right\} du, \quad (16)$$

where $J_0(\cdot)$ denotes the zeroth-order Bessel function of the first kind [33, Eq. (8.411/1)]. Eventually, substituting (16) into (14) and changing the order of the resultant integrals, and then performing algebraic manipulations using $(\partial/\partial p)\,J_0\left(2\sqrt{pu}\right) = -\sqrt{u/p}\,J_1\left(2\sqrt{pu}\right)$, we have

$$\mathcal{P}_{AUP} = c - \frac{n}{2}\left\{c + \int_0^\infty \frac{(-1)^n c}{\Gamma(b)} \sum_{\ell=1}^{L} \mu_\ell \mathcal{I}(u) \times \right.$$
$$\left.\left\{\mathcal{M}_{1/\widehat{\gamma}_{1,\ell}}(u)\frac{\partial}{\partial u}\mathcal{M}_{1/\gamma_{2,\ell}}(u) + \mathcal{M}_{1/\gamma_{2,\ell}}(u)\frac{\partial}{\partial u}\mathcal{M}_{1/\widehat{\gamma}_{1,\ell}}(u)\right\} du\right\}, \quad (17)$$

where the function $\mathcal{I}(u)$ can be obtained in closed-form by means of utilizing [31, Eq. (2.24.4/1)] and [31, Eqs. (8.2.2/14) and (8.2.2/15)] as

$$\mathcal{I}(u) = \int_0^\infty \sqrt{\frac{u}{p}}\, J_1\left(2\sqrt{pu}\right) \mathrm{G}_{n+1,2}^{1,n}\left[\frac{a}{p}\,\middle|\,\begin{matrix}\Lambda_1^{(n)}, 1\\ b, 0\end{matrix}\right] dp = \mathrm{G}_{n,3}^{1,n}\left[au\,\middle|\,\begin{matrix}\Lambda_1^{(n)}\\ b, 0, 0\end{matrix}\right]. \quad (18)$$

Then, substituting (18) into (19) results in

$$\mathcal{P}_{AUP} = c - \frac{n}{2}\left\{c + \int_0^\infty \frac{(-1)^n c}{\Gamma(b)} \sum_{\ell=1}^{L} \mu_\ell \mathrm{G}_{n,3}^{1,n}\left[au\,\middle|\,\begin{matrix}\Lambda_1^{(n)}\\ b, 0, 0\end{matrix}\right] \times \right.$$
$$\left.\left\{\mathcal{M}_{1/\widehat{\gamma}_{1,\ell}}(u)\frac{\partial}{\partial u}\mathcal{M}_{1/\gamma_{2,\ell}}(u) + \mathcal{M}_{1/\gamma_{2,\ell}}(u)\frac{\partial}{\partial u}\mathcal{M}_{1/\widehat{\gamma}_{1,\ell}}(u)\right\} du\right\}, \quad (19)$$

where the terms outside of the integration can be easily put inside the integration by using the well-known fact that

$$\int_0^\infty \left\{ \mathcal{M}_{1/\widehat{\gamma}_{1,\ell}}(u) \frac{\partial}{\partial u} \mathcal{M}_{1/\gamma_{2,\ell}}(u) + \mathcal{M}_{1/\gamma_{2,\ell}}(u) \frac{\partial}{\partial u} \mathcal{M}_{1/\widehat{\gamma}_{1,\ell}}(u) \right\} du = -1. \quad (20)$$

Finally, applying (20) on (19) readily results in (12), which proves Theorem 1. ∎

It is both useful and informative to notice that the exact AUP $\mathcal{P}_{AUP}$ presented in Theorem 1 is straightforward to utilize depending on the exact closed-form expressions for the reciprocal MGFs $\mathcal{M}_{1/\widehat{\gamma}_{1,\ell}}(p)$ and $\mathcal{M}_{2/\gamma_{2,\ell}}(p)$. However, both $\widehat{\gamma}_{1,\ell}$ and $\gamma_{2,\ell}$ are two composite distributions. Specifically, the instantaneous SNR $\widehat{\gamma}_{1,\ell}$ is given by $\widehat{\gamma}_{1,\ell} = \widehat{\mathcal{S}}_{1,\ell} \mathcal{G}_{1,\ell}$, then on the condition that the SSI knowledge, i.e., $\widehat{\mathcal{S}}_{1,\ell}$ is available, the reciprocal MGF $\mathcal{M}_{1/\widehat{\gamma}_{1,\ell}}(p) = \int_0^\infty \mathbb{E}\left[e^{-p/\widehat{\gamma}_{1,\ell}} \middle| \widehat{\mathcal{S}}_{1,\ell} = s\right] p_{\widehat{\mathcal{S}}_{1,\ell}}(s) \, ds$ is written as

$$\mathcal{M}_{1/\widehat{\gamma}_{1,\ell}}(p) = \int_0^\infty \mathcal{M}_{1/\mathcal{G}_{1,\ell}}(p/s) \, p_{\widehat{\mathcal{S}}_{1,\ell}}(s) \, ds, \quad (21)$$

where $\mathcal{M}_{1/\mathcal{G}_{1,\ell}}(p)$ is the reciprocal MGF of the multipath fast fading that the $\ell$th RT is exposed to. In turn as explained before, $\widehat{\mathcal{S}}_{1,\ell} \triangleq \max_{k \in \{1,2,\ldots,L\}} \mathcal{S}_{1,k}$ denotes the shadowing power of the first hop conditioned on that the $\ell$th RT is selected. Its PDF $p_{\widehat{\mathcal{S}}_{1,\ell}}(s)$ is given by

$$p_{\widehat{\mathcal{S}}_{1,\ell}}(s) = \sum_{l=1}^L p_{\mathcal{S}_{1,l}}(s) \prod_{\substack{k=1 \\ k \neq l}}^L P_{\mathcal{S}_{1,k}}(s). \quad (22)$$

where for $\ell \in \{1,2,\ldots,L\}$ $p_{\mathcal{S}_{1,\ell}}(s)$ and $P_{\mathcal{S}_{1,\ell}}(s)$ are defined before. Note that referring [27], $\mathcal{M}_{1/\mathcal{G}_{1,\ell}}(p)$ results in closed-form expressions for the fading distributions commonly used in the literature. As a result of that, the derivative of $\mathcal{M}_{1/\mathcal{G}_{1,\ell}}(p)$ is also available in closed-form in the literature. Thus, the complexity on the computation of the reciprocal MGF $\mathcal{M}_{1/\widehat{\gamma}_{1,\ell}}(p)$ is simplified by all means. As a step forward, using Gauss-Chebyshev quadrature (GCQ) formula [30, Eq.(25.4.39)], the reciprocal MGF $\mathcal{M}_{1/\widehat{\gamma}_{1,\ell}}(p)$ can be written as

$$\mathcal{M}_{1/\widehat{\gamma}_{1,\ell}}(p) = \sum_{n=1}^N \eta_{n,\ell} \mathcal{M}_{1/\mathcal{G}_{1,\ell}}(p/s_n), \quad (23)$$

where the coefficient $\eta_{n,\ell}$ is given by

$$\eta_{n,\ell} = w_n \sum_{l=1}^L p_{\mathcal{S}_{1,l}}(s_n) \prod_{\substack{k=1 \\ k \neq l}}^L P_{\mathcal{S}_{1,k}}(s_n). \quad (24)$$

Furthermore in (23) and (24), the coefficients $s_n$ and $w_n$ are defined in [35, Eqs. (22) and (23)], respectively, and where the truncation index $N$ could be sufficiently chosen for an enough accurate result. In addition, conditioned on that the $\ell$th RT is selected, the instantaneous SNR of the second hop, i.e., $\gamma_{2,\ell}$ is given by $\gamma_{2,\ell} = \mathcal{S}_{2,\ell} \mathcal{G}_{2,\ell}$. Then, the reciprocal MGF $\mathcal{M}_{1/\gamma_{2,\ell}}(p) = \mathbb{E}[\exp(-p/\gamma_{2,\ell})]$, $\Re\{p\} \in \mathbb{R}^+$ is obtained as

$$\mathcal{M}_{1/\gamma_{2,\ell}}(p) = \int_0^\infty \mathcal{M}_{1/\mathcal{G}_{2,\ell}}(p/s) \, p_{\mathcal{S}_{2,\ell}}(s) \, ds \tag{25}$$

where $\mathcal{M}_{2/\mathcal{G}_{2,\ell}}(p)$ is the reciprocal MGF of the multipath fast fading that the $\ell$th RT is exposed to. In the special case of that the RTs $R_2, R_2, \ldots, R_L$ are much more closer to the DT, which create a cluster of RTs whose channels towards to the DT $D$ have very negligible power fluctuation as compared with their average powers. In other words, $\text{Var}\left[\mathcal{S}_{2,\ell}\right] \approx 0$ for all RTs $\ell \in \{1, 2, \ldots, L\}$. Therefore, the distribution of the shadowing $\mathcal{S}_{2,\ell}$ turns into Dirac's distribution and its corresponding PDF $p_{\mathcal{S}_{2,\ell}}(s)$ definitely becomes $p_{\mathcal{S}_{2,\ell}}(s) = \delta(s - \Omega_{2,\ell})$. Immediately substituting this PDF into (25), $\mathcal{M}_{1/\gamma_{2,\ell}}(p)$ results in

$$\mathcal{M}_{1/\gamma_{2,\ell}}(p) = \mathcal{M}_{1/\mathcal{G}_{2,\ell}}(p/\Omega_{2,\ell}). \tag{26}$$

Finally, note that (23), (24), (25) and (26) can be efficiently computed for a variety of commonly used fading distributions such as Rayleigh, lognormal, Weibull, Nakagami-$m$, generalized Nakagami-$m$, generalized-K and EGK distributions. Consequently, substituting them into (13), the exact AUP $\mathcal{P}_{AUP}$ is easily and efficiently computed.

## B. Special Cases of the Auxiliary Function $\mathcal{Z}_{a,b,c}^n(u)$

Let us consider in this section the special cases of the auxiliary function $\mathcal{Z}_{a,b,c}^n(u)$ in order to check and show analytical simplicity, correctness and accuracy:

**Special Case 1** (Auxiliary Functions for the Bit Error Probabilities of the Coherent Modulation Schemes)**.** Setting $n = 1$, $c = 1$ and $b = 1/2$ in (12) in order to find the auxiliary functions for the bit error probabilities of coherent BFSK and BPSK modulation schemes, the auxiliary function $\mathcal{Z}_{a,b}^n(u)$ simplifies to

$$\mathcal{Z}_{a,1/2,1}^1(u) = \frac{1}{2} - \frac{1}{\pi}\text{Si}\left(2\sqrt{au}\right), \tag{27}$$

by means of using [31, Eq. (8.4.12/1)], where $\text{Si}(\cdot)$ is the sine integral function defined as $\text{Si}(z) = \int_0^z u^{-1}\sin(u) \, du$ [30, Eq. (5.2.1)], and where the parameter $a = 1/2$ for orthogonal coherent FSK (BFSK) [36, Eq. (8.43)], and $a = 1$ for antipodal coherent PSK (BPSK) [36, Eq. (8.19)]. Inserting (27) into (12), the ABEP of the partial relay selection is obtained for BFSK and BPSK modulation schemes. □

**Special Case 2** (Auxiliary Functions for the Bit Error Probabilities of the Non-Coherent Modulation Schemes)**.** In the special case of non-coherent modulation schemes, substituting $n = 1$, $c = 1$ and $b = 1$ (13) simplifies the auxiliary function by means of using [31, Eq. (8.4.19/1)], that is

$$\mathcal{Z}_{a,1,1}^1(u) = \frac{1}{2}J_0\left(2\sqrt{au}\right), \tag{28}$$

where the parameter $a = 1/2$ for orthogonal non-coherent FSK (NCFSK) [36, Eq. (8.69)] and $a = 1$ for antipodal differentially coherent BPSK (BDPSK) [36, Eq. (8.85)]. Eventually by substituting (28) into (12), the ABEP of the partial relay selection is easily obtained for non-coherent NCFSK and BDPSK modulation schemes. □

**Special Case 3** (Auxiliary Functions for Bit Error Probabilities of the Wojnar's Representation). For conditional BEP $P_{BEP}(\gamma_{end})$ proposed by Wojnar in [32, Eq. (13)], setting $n = 1$ and $c = 1$ in (12), the auxiliary function expressly simplifies to

$$\mathcal{Z}^1_{a,b,1}(u) = \frac{1}{2} - \frac{1}{2\Gamma(b)} \mathrm{G}^{1,1}_{1,3}\left[au \left|\begin{array}{c} 1 \\ b,0,0 \end{array}\right.\right]. \tag{29}$$

After performing some algebraic manipulations either by means of using [37, Property 2.11] or by recognizing that the CDF of a Fox's H distribution can be written in two different forms given in [38, Eq. (4.17)] and [38, Eq. (4.19)], (29) simplifies to

$$\mathcal{Z}^1_{a,b,1}(u) = \frac{1}{2\Gamma(b)} \mathrm{G}^{2,0}_{1,3}\left[au \left|\begin{array}{c} 1 \\ b,0,0 \end{array}\right.\right] \tag{30}$$

which is in perfect agrement with the well-known result given in [39, Eq. (14)]. In more details using [31, Eq. (8.4.12/1) and (8.4.19/1)], substituting $b = 1/2$ and $b = 1$ into (30) result in (27) and (28), respectively. Finally, inserting (30) into (12), the ABEP of the partial relay selection can be readily obtained for all coherent and non-coherent binary modulation schemes.

**Special Case 4** (Auxiliary Function for Ergodic Capacity). Substituting $a = 1$, $b = 1$, $c = W/\log(2)$ and $n = 2$ into (12), the auxiliary function for the EC analysis is easily obtained as

$$\mathcal{Z}^2_{1,1,\frac{W}{\log(2)}}(u) = \frac{W}{\log(2)} \mathrm{G}^{1,2}_{2,3}\left[u \left|\begin{array}{c} 1,1 \\ 1,0,0 \end{array}\right.\right]. \tag{31}$$

Note that utilizing [40, Eq. (07.34.03.0475.01)] and [40, Eq. (06.06.03.0003.01)] together, (31) further simplifies into [41, Eq. (20)], that is

$$\mathcal{Z}^2_{1,1,\frac{W}{\log(2)}}(u) = \frac{W}{\log(2)}\left(\log(u) - \mathrm{Ei}(-u) + \mathbf{C}\right), \tag{32}$$

where $\mathrm{Ei}(\cdot)$ is the exponential integral function [33, Eq. (8.211/1)] and $\mathbf{C}$ is the Euler-Mascheroni constant (also called Euler's constant) [33, Eq.(8.367/1)]. Eventually, inserting (32) into (12), the EC of the partial relay selection can be readily obtained. □

**Special Case 5** (Auxiliary Function for the Moments-Generating Function). Substituting $a = p$, $b = 1$, $c = 2$ and $n = 1$ into (12), the auxiliary function for the MGF analysis is easily obtained as

$$\mathcal{Z}_{p,1,2}^{1}(u) = 1 - \mathrm{G}_{1,3}^{1,1}\left[pu \,\middle|\, \begin{matrix} 1 \\ 1,0,0 \end{matrix}\right]. \tag{33}$$

Utilizing [40, Eq. (07.34.03.0285.01)], [31, Eq. (7.14.2/84)] and [30, Eq. (9.6.3)], (33) further simplifies into the transform function of [42, Eq. (13)], that is

$$\mathcal{Z}_{p,1,2}^{1}(u) = J_0\left(2\sqrt{pu}\right). \tag{34}$$

Finally substituting (34) into (12) yields (16) as expected. □

Consequently, the mathematical formalism in this subsection shows how to treat the ABEP analysis of coherent and non-coherent binary modulation schemes, EC analysis and MGF-based statistical analysis of the overall instantaneous SNR simultaneously in parallel. As such, these different analyzes could be evidently viewed as a single problem [29] for a variety of fading distributions.

### C. The Moments of the Overall Instantaneous SNR

It is worthy to mention that the moments of the overall instantaneous SNR $\gamma_{end}$ is another important tool to characterize the probability distribution such that it can be used to obtain the higher-order amount-of-fading (AF) defined as [43, Eq. (3)]

$$AF_{\gamma_{end}}^{(k)} = \frac{\mathbb{E}[\gamma_{end}^k]}{\mathbb{E}[\gamma_{end}]^k} - 1, \tag{35}$$

where the $k$th moment of the overall instantaneous SNR $\gamma_{end}$ can be obtained with the aid of $\mathbb{E}[\gamma_{end}^k] = (\partial/\partial p)^k \mathcal{M}_{\gamma_{end}}(p)\big|_{p\to 0}$, where $\mathcal{M}_{\gamma_{end}}(p)$ is given in (16). Within this concept, (34) is very useful. Accordingly, the $k$th derivative of (34) with respect to $p$ at $p=0$ is derived by means of performing some algebraic manipulations using [33, Eq. (8.411/1) and (8.411/2)], that is

$$(-1)^k \frac{\partial^k}{\partial p^k}\mathcal{Z}_{p,1,2}^{1}(u)\bigg|_{p\to 0} = (-1)^k \frac{\partial^k}{\partial p^k} J_0\left(2\sqrt{pu}\right)\bigg|_{p\to 0} = \frac{u^k}{k!}. \tag{36}$$

Utilizing (36), the $k$th moment of the overall instantaneous SNR $\gamma_{end}$ is easily obtained as

$$\mathbb{E}[\gamma_{end}^k] = -\int_0^\infty \frac{u^k}{k!}\sum_{\ell=1}^{L}\mu_\ell\left\{\mathcal{M}_{1/\widehat{\gamma}_{1,\ell}}(u)\frac{\partial}{\partial u}\mathcal{M}_{1/\gamma_{2,\ell}}(u) + \mathcal{M}_{1/\gamma_{2,\ell}}(u)\frac{\partial}{\partial u}\mathcal{M}_{1/\widehat{\gamma}_{1,\ell}}(u)\right\}du. \tag{37}$$

Then with aid of (37), the higher-order AF can be readily obtained for a variety of fading distributions. In addition, (37) can be utilized in either asymptotic performance analysis [43], [44] or approximation of the outage probability using Laguerre moments [34].

Despite the fact that the novel and generic technique represented in this section is easy to use for a variety of generalized fading distributions, let us consider in the next section some numerical and simulation examples achieved for the EGK fading environments in order to check its analytical simplicity and accuracy.

## IV. Applications in Generalized Composite Fading Environments

We consider the partial-RSP based on the SSI of the fading channels between the ST $S$ and the RTs $R_1, R_2, \ldots, R_L$, as depicted in Fig. 1 for composite fading environments. However, even if the results obtained in the previous sections are in general applicable to many generalized composite fading environments, our results in this section have been intentionally carried for the EGK fading distribution in order to check numerically the correctness and accuracy of the presented mathematical formalism. In the following subsection, the EGK distribution model is explained and its some important properties are accentuated.

### A. Extended Generalized-K Fading Environments

The EGK distribution is a generalized distribution that provides a unified model to statistically characterize the envelope statistics of wireless communication channels as explained in [26], [27]. More precisely, as listed in both [26, Table I] and [27, Table I], several distributions such as Rayleigh, lognormal, Weibull, Nakagami-$m$, generalized Nakagami-$m$, generalized-K are special or limiting cases of the EGK distribution. Referring to the partial-RSP depicted in Fig. 1, for all hops $k \in \{1, 2\}$ and RTs $\ell \in \{1, 2, \ldots, L\}$, the instantaneous SNR $\gamma_{k,\ell}$ is assumed to follow the EGK distribution whose PDF can be given in [26, Eq.(3)], [27, Eq.(5)] by

$$p_{\gamma_{k,\ell}}(\gamma) = \frac{\xi_{k,\ell} \left(\frac{\phi_{k,\ell}\varphi_{k,\ell}}{\Omega_{k,\ell}}\right)^{m_{k,\ell}\xi_{k,\ell}}}{\Gamma(m_{k,\ell})\Gamma(n_{k,\ell})} \gamma^{m_{k,\ell}\xi_{k,\ell}-1} \Gamma\left(n_{k,\ell} - m_{k,\ell}\frac{\xi_{k,\ell}}{\zeta_{k,\ell}}, 0, \left(\frac{\phi_{k,\ell}\varphi_{k,\ell}}{\Omega_{k,\ell}}\right)^{m_{k,\ell}\xi_{k,\ell}} \gamma^{\xi_{k,\ell}}, \frac{\xi_{k,\ell}}{\zeta_{k,\ell}}\right), \qquad (38)$$

where $m_{k,\ell}$ ($0.5 \leq m_{k,\ell} < \infty$) and $\xi_{k,\ell}$ ($0 \leq \xi_{k,\ell} < \infty$) represent the fading figure (diversity severity / order) and the fading shaping factor, respectively, while $n_{k,\ell}$ ($0.5 \leq n_{k,\ell} < \infty$) and $\zeta_{k,\ell}$ ($0 \leq \zeta_{k,\ell} < \infty$) represent the shadowing severity and the shadowing shaping factor (inhomogeneity), respectively. In addition, the parameters $\phi_{k,\ell}$ and $\varphi_{k,\ell}$ are defined as $\phi_{k,\ell} = \Gamma(m_{k,\ell} + 1/\xi_{k,\ell})/\Gamma(m_{k,\ell})$ and $\varphi_{k,\ell} = \Gamma(n_{k,\ell} + 1/\zeta_{k,\ell})/\Gamma(n_{k,\ell})$, respectively. Furthermore, in (38), $\Gamma(\cdot, \cdot, \cdot, \cdot)$ is the extended incomplete Gamma function defined as $\Gamma(\alpha, x, b, \beta) = \int_x^\infty r^{\alpha-1} \exp\left(-r - br^{-\beta}\right) dr$, where $\alpha, \beta, b \in \mathbb{C}$ and $x \in \mathbb{R}^+$ [45, Eq. (6.2)].

In addition, its MGF, i.e., $\mathcal{M}_{\gamma_{k,\ell}}(p) \equiv \mathbb{E}\left[\exp(-p\gamma_{k,\ell})\right]$ is given in [27, Eq. (15)]. Then, using [34, Theorem 1], its reciprocal MGF $\mathcal{M}_{1/\gamma_{k,\ell}}(p) \equiv \mathbb{E}\left[\exp(-p/\gamma_{k,\ell})\right]$ is obtained as

$$\mathcal{M}_{1/\gamma_{k,\ell}}(p) = \frac{1}{\Gamma(m_{k,\ell})\Gamma(n_{k,\ell})} \text{H}_{0,3}^{3,0}\left[\frac{\phi_{k,\ell}\varphi_{k,\ell}}{\Omega_{k,\ell}}p \left| \begin{array}{c} --- \\ (0,1), (m_{k,\ell}, 1/\xi_{k,\ell}), (n_{k,\ell}, 1/\zeta_{k,\ell}) \end{array} \right. \right], \quad (39)$$

where $\text{H}_{p,q}^{m,n}[\cdot]$ denotes the Fox's H function[1,2] defined in [31, Eq.(8.3.1/1)], [37, Eq.(1.1.1)], and where $---$ means that the coefficients are absent.

As mentioned before, the instantaneous SNR $\gamma_{k,\ell}$ is defined as $\gamma_{k,\ell} = \mathcal{S}_{k,\ell}\mathcal{G}_{k,\ell}$. When $\gamma_{k,\ell}$ follows the EGK distribution, then its multipath fast fading part, i.e., $\mathcal{G}_{k,\ell}$ follows the well-known generalized gamma (GG) distribution with the PDF $p_{\mathcal{G}_{k,\ell}}(g) \equiv \mathbb{E}[\delta(g - \gamma_{k,\ell})|\mathcal{S}_{k,\ell} = 1]$, $g \in \mathbb{R}^+$, where assumed without loss of generality that the multipath fast fading $\mathcal{G}_{k,\ell}$ has unit power, i.e., $\mathbb{E}[\mathcal{G}_{k,\ell}] = 1$. Then, the PDF $p_{\mathcal{G}_{k,\ell}}(g)$, $g \in \mathbb{R}^+$ is explicitly given by

$$p_{\mathcal{G}_{k,\ell}}(g) = \frac{\xi_{k,\ell}\,\phi_{k,\ell}}{\Gamma(m_{k,\ell})}(\phi_{k,\ell}g)^{m_{k,\ell}\xi_{k,\ell}-1} e^{-(\phi_{k,\ell}g)^{\xi_{k,\ell}}}. \quad (40)$$

Note that as indicated in [39], the reciprocal MGF $\mathcal{M}_{1/\mathcal{G}_{k,\ell}}(p) = \mathbb{E}[\exp(-p/\mathcal{G}_{k,\ell})]$ and its first-order derivative $(\partial/\partial p)\mathcal{M}_{1/\mathcal{G}_{k,\ell}}(p) = -\mathbb{E}[\exp(-p/\mathcal{G}_{k,\ell})/\mathcal{G}_{k,\ell}]$ are required to obtain computationally efficient analytical results. With the aid of the result in [39, Eq. (18)], the reciprocal MGF of the multipath fast fading $\mathcal{G}_{k,\ell}$ is written as

$$\mathcal{M}_{1/\mathcal{G}_{k,\ell}}(p) = \frac{1}{\Gamma(m_{k,\ell})}\Gamma\left(m_{k,\ell}, 0, \phi_{k,\ell}p, 1/\xi_{k,\ell}\right). \quad (41)$$

Accordingly using [45, Eq.(6.22)] and [31, Eq.(8.2.2/30)] together, its first-order derivative can be written as [39, Eq. (19)]

$$\frac{\partial}{\partial p}\mathcal{M}_{1/\mathcal{G}_{k,\ell}}(p) = -\frac{\phi_{k,\ell}}{\Gamma(m_{k,\ell})}\Gamma\left(m_\ell - 1/\xi_\ell, 0, \phi_{k,\ell}p, 1/\xi_{k,\ell}\right). \quad (42)$$

Note that the average power of $\gamma_{k,\ell}$, i.e., $\mathbb{E}[\gamma_{k,\ell}] = \mathcal{S}_{k,\ell}$ changes slowly, so without loss of generality $\mathcal{S}_{k,\ell}$ can be assumed a constant for a short period of time but has a distribution over a long period of time. Explicitly, its distribution explicitly follows the PDF given by $p_{\mathcal{S}_{k,\ell}}(s) \equiv \mathbb{E}[\delta(s - \mathbb{E}[\gamma_{k,\ell}])] = \mathbb{E}[\delta(s - \mathcal{S}_{k,\ell})]$, $s \in \mathbb{R}^+$ given by

$$p_{\mathcal{S}_{k,\ell}}(s) = \frac{\zeta_{k,\ell}\varphi_{k,\ell}}{\Gamma(n_{k,\ell})\Omega_{k,\ell}}\left(\frac{\varphi_{k,\ell}}{\Omega_{k,\ell}}s\right)^{n_{k,\ell}\zeta_{k,\ell}-1} e^{-\left(\frac{\varphi_{k,\ell}}{\Omega_{k,\ell}}s\right)^{\zeta_{k,\ell}}}, \quad (43)$$

where $\mathbb{E}[\gamma_{k,\ell}] = \mathbb{E}[\mathcal{S}_{k,\ell}] = \Omega_{k,\ell}$ for a long period of time. As mentioned in the previous subsection that the selection of the RT will be achieved based on the SSI information of the channels / hops. The CDF of $\mathcal{S}_{k,\ell}$, i.e., $P_{\mathcal{S}_{k,\ell}}(s) = \mathbb{E}[\theta(s - \mathbb{E}[\gamma_{k,\ell}])] = \mathbb{E}[\theta(s - \mathcal{S}_{k,\ell})]$, $s \in \mathbb{R}^+$ is therefore required and given by

$$P_{\mathcal{S}_{k,\ell}}(s) = \frac{1}{\Gamma(n_{k,\ell})} \gamma\left(n_{k,\ell}, \left(s\,\varphi_{k,\ell}/\Omega_{k,\ell}\right)^{\zeta_{k,\ell}}\right), \tag{44}$$

where $\gamma(\cdot,\cdot)$ is the upper incomplete Gamma function [33, Eqs. (8.350/1)].

---

[1] Using [31, Eq.(8.3.2/22)], the Fox's H function can be represented in terms of the Meijer's G function [31, Eq.(8.2.1/1)] which is a built-in function in the most popular mathematical software packages such as MATHEMATICA®

[2] The authors gives in the appendix of [46] an efficient implementation of the Fox's H function in MATHEMATICA®

### B. Unified Performance Analysis in Extended Generalized-K Fading Environments

Regarding the numerical and simulation results, we considered the EGK fading environment for different number of RTs, i.e., $L \in \{1,2,3,5\}$ with the statistical channel parameters $m_{k,\ell}$, $\xi_{k,\ell}$, $n_{k,\ell}$, $\zeta_{k,\ell}$ and $\Omega_{k,\ell}$ for all hops $k \in \{1,2\}$ and all RTs $\ell \in \{1,2,\ldots,L\}$ as shown in Table I. In order to obtain the AUP $\mathcal{P}_{AUP}$ of the SSI-based partial relay selection in EGK fading environments, Theorem 1 introduces an MGF-based approach that requires only the reciprocal MGFs $\mathcal{M}_{1/\hat{\gamma}_{1,\ell}}(p)$ and $\mathcal{M}_{1/\gamma_{2,\ell}}(p)$ in closed-form. As such, the reciprocal MGF $\mathcal{M}_{1/\hat{\gamma}_{1,\ell}}(p)$ can be obtained by substituting (42) and (44) into (23), that is

$$\mathcal{M}_{1/\hat{\gamma}_{1,\ell}}(p) = \sum_{n=1}^{N} \frac{\eta_{n,\ell}}{\Gamma(m_{1,\ell})} \Gamma\left(m_{1,\ell}, 0, \frac{\phi_{1,\ell}\, p}{s_n}, \frac{1}{\xi_{1,\ell}}\right). \tag{45}$$

where the coefficients $\{\eta_{n,\ell}\}$ are easily computed substituting both (43) and (44) into (24). Accordingly using (42), its derivative can be easily given as

$$\frac{\partial}{\partial p}\mathcal{M}_{1/\hat{\gamma}_{1,\ell}}(p) = -\sum_{n=1}^{N} \frac{\eta_{n,\ell}\,\phi_{1,\ell}}{\Gamma(m_{1,\ell})\, s_n} \Gamma\left(m_{1,\ell}-\frac{1}{\xi_{1,\ell}}, 0, \frac{\phi_{1,\ell}\, p}{s_n}, \frac{1}{\xi_{1,\ell}}\right). \tag{46}$$

In addition, conditioned on that the $\ell$th RT has been selected, the reciprocal MGF of the second hop, $\mathcal{M}_{1/\gamma_{2,\ell}}(p)$ can be obtained using (39), that is

$$\mathcal{M}_{1/\gamma_{2,\ell}}(p) = \frac{1}{\Gamma(m_{2,\ell})\Gamma(n_{2,\ell})} H^{3,0}_{0,3}\left[\frac{\phi_{2,\ell}\,\varphi_{2,\ell}}{\Omega_{2,\ell}} p \,\bigg|\, \begin{matrix}---\\(0,1), (m_{2,\ell}, 1/\xi_{2,\ell}), (n_{2,\ell}, 1/\zeta_{2,\ell})\end{matrix}\right]. \tag{47}$$

As mentioned before, the RTs $R_1, R_2, \ldots, R_L$ are, without loss of generality, assumed to be very close to the DT $D$, accordingly creating such a RT cluster that the transmission channels from RTs to the DT $D$ are not subjected to disruptive shadowing effects as explained in Section **??**. Evidently in such a case, the shadowing conditions certainly disappear since the shadowing figures $\{n_{2,\ell}\}$ explicitly approach to infinity when the RTs $R_1, R_2, \ldots, R_L$ get closer to the DT $D$. Then refering (41), the reciprocal MGF of the second hop, $\mathcal{M}_{1/\gamma_{2,\ell}}(p)$ reduces to $\mathcal{M}_{1/\mathcal{G}_{2,\ell}}(p/\Omega_{2,\ell})$, that is

$$\mathcal{M}_{1/\gamma_{2,\ell}}(p) = \frac{1}{\Gamma(m_{2,\ell})} \Gamma\left(m_{2,\ell}, 0, \frac{\phi_{2,\ell}\, p}{\Omega_{2,\ell}}, \frac{1}{\xi_{2,\ell}}\right). \tag{48}$$

Accordingly using (42), its derivative can be easily given as

$$\frac{\partial}{\partial p}\mathcal{M}_{1/\gamma_{2,\ell}}(p) = -\frac{\phi_{2,\ell}}{\Gamma(m_{2,\ell})\,\Omega_{2,\ell}} \Gamma\left(m_{2,\ell}-\frac{1}{\xi_{2,\ell}}, 0, \frac{\phi_{2,\ell}\, p}{\Omega_{2,\ell}}, \frac{1}{\xi_{2,\ell}}\right). \tag{49}$$

Eventually, substituting the set of Equations (45), (46), (48) and (49) into (12), the AUP $\mathcal{P}_{AUP}$ of the SSI-based partial-RSP in dual-hop AF system over EGK fading environment can be readily obtained as

$$\mathcal{P}_{AUP} = \int_0^\infty \mathcal{Z}_{a,b,c}^n(u) \sum_{n=1}^N \sum_{\ell=1}^L \frac{\eta_{n,\ell}\,\mu_\ell}{\Gamma(m_{1,\ell})\Gamma(m_{2,\ell})} \Bigg\{$$
$$\frac{\phi_{1,\ell}}{s_n}\Gamma\left(m_{1,\ell}-\frac{1}{\xi_{1,\ell}},0,\frac{\phi_{1,\ell}\,u}{s_n},\frac{1}{\xi_{1,\ell}}\right)\Gamma\left(m_{2,\ell},0,\frac{\phi_{2,\ell}\,u}{\Omega_{2,\ell}},\frac{1}{\xi_{2,\ell}}\right)+$$
$$\frac{\phi_{2,\ell}}{\Omega_{2,\ell}}\Gamma\left(m_{1,\ell},0,\frac{\phi_{1,\ell}\,u}{s_n},\frac{1}{\xi_{1,\ell}}\right)\Gamma\left(m_{2,\ell}-\frac{1}{\xi_{2,\ell}},0,\frac{\phi_{2,\ell}\,u}{\Omega_{2,\ell}},\frac{1}{\xi_{2,\ell}}\right)\Bigg\}du. \quad (50)$$

Note that the coefficients $\mu_\ell$ in (50), which is the probability that the $\ell$th RT is selected, can be readily computed by substituting both (43) and (44) into (3). For the well-known generalized-K (GK) fading environment (i.e., for the special case of the EGK fading environment with the fading and shadowing shaping parameters $\xi_{k,\ell} = \zeta_{k,\ell} = 1$ for all $k \in \{1,2\}$ and $\ell \in \{1,2,\ldots,L\}$), the unified result given in (50) can be more simplified to

$$\mathcal{P}_{AUP} = \int_0^\infty \mathcal{Z}_{a,b,c}^n(u) \sum_{n=1}^N \sum_{\ell=1}^L \frac{4\sqrt{u^{m_{1,\ell}+m_{2,\ell}-1}}\,\eta_{n,\ell}\,\mu_\ell}{\Gamma(m_{1,\ell})\Gamma(m_{2,\ell})} \sqrt{\frac{(m_{1,\ell}/s_n)^{m_{1,\ell}}}{(\Omega_{2,\ell}/m_{2,\ell})^{m_{2,\ell}}}} \Bigg\{$$
$$\sqrt{\frac{m_{1,\ell}}{s_n}} K_{m_{1,\ell}-1}\left(2\sqrt{\frac{m_{1,\ell}\,u}{s_n}}\right) K_{m_{2,\ell}}\left(2\sqrt{\frac{m_{2,\ell}\,u}{\Omega_{2,\ell}}}\right)+$$
$$\sqrt{\frac{m_{2,\ell}}{\Omega_{2,\ell}}} K_{m_{1,\ell}}\left(2\sqrt{\frac{m_{1,\ell}\,u}{s_n}}\right) K_{m_{2,\ell}-1}\left(2\sqrt{\frac{m_{2,\ell}\,u}{\Omega_{2,\ell}}}\right)\Bigg\}du. \quad (51)$$

by means of using [45, Eq. (6.41)], i.e.,

$$\Gamma(m,0,b,1) = 2b^{m/2} K_m\left(2\sqrt{b}\right), \quad \Re\{b\} \in \mathbb{R}^+, \quad (52)$$

where $K_m(\cdot)$ denotes the $m$th order modified Bessel function of the second kind [30, Eq. (9.6.2)]. Note that, under the further assumptions that the shadowing distributions $\{\mathcal{S}_{1,\ell}\}$ are identically distributed (i.e., $n_{1,\ell} = n$, $\zeta_{1,\ell} = \zeta$ and $\Omega_{1,\ell} = \Omega$ for all $\ell \in \{1,2,\ldots,L\}$ such that $\mathcal{S}_{1,\ell} \to \mathcal{S}_1$ for all $\ell \in \{1,2,\ldots,L\}$), each RT will have the same probability to be selected, i.e., the probability of the $\ell$th RT is selected certainly becomes

$$\mu_\ell = \int_0^\infty p_{\mathcal{S}_1}(s)\left[P_{\mathcal{S}_1}(s)\right]^{L-1} ds = \frac{1}{L}, \quad (53)$$

for all $\ell \in \{1,2,\ldots,L\}$. In such circumstances, when the shadowing conditions get better (i.e., the disruptive effects of the shadowing conditions on the received power decreases) as the shadowing figure surely goes to infinity ($n \to \infty$). Then, the SSI-based partial-RSP turns into the simplest round-robin (RR)- based partial relay selection tournaments in which each RT is periodically chosen for signal transmission. Although it has worse performance than the other RSPs, it provides quite an ABEP upper-bound (i.e., worst-case) performance that the SSI-based partial relay selection can achieve. In such a case, (50) simplifies as expected to

$$\mathcal{P}_{AUP} = \int_0^\infty \mathcal{Z}_{a,b}^n(u) \sum_{\ell=1}^L \frac{1/L}{\Gamma(m_{1,\ell})\Gamma(m_{2,\ell})} \left\{ \frac{\phi_{1,\ell}}{\Omega} \Gamma\left(m_{1,\ell} - \frac{1}{\xi_{1,\ell}}, 0, \frac{\phi_{1,\ell} u}{\Omega}, \frac{1}{\xi_{1,\ell}}\right) \Gamma\left(m_{2,\ell}, 0, \frac{\phi_{2,\ell} u}{\Omega_{2,\ell}}, \frac{1}{\xi_{2,\ell}}\right) + \right.$$
$$\left. \frac{\phi_{2,\ell}}{\Omega_{2,\ell}} \Gamma\left(m_{1,\ell}, 0, \frac{\phi_{1,\ell} u}{\Omega}, \frac{1}{\xi_{1,\ell}}\right) \Gamma\left(m_{2,\ell} - \frac{1}{\xi_{2,\ell}}, 0, \frac{\phi_{2,\ell} u}{\Omega_{2,\ell}}, \frac{1}{\xi_{2,\ell}}\right) \right\} du. \quad (54)$$

On the other hand, when the shadowing distributions $\{\mathcal{S}_{1,\ell}\}$ are not identically distributed (i.e., $\Omega_{1,\ell} \neq \Omega_{1,k}$, $n_{1,\ell} \neq n_{1,k}$ and $\zeta_{1,\ell} \neq \zeta_{1,k}$ for $\ell \neq k$), and when the shadowing figures goes to infinity (for all $\ell \in \{1, 2, \ldots, L\}$, $n_{1,\ell} \to \infty$), the CE will definitely select the RT having the highest AP, that is

$$\hat{\ell} = \arg \max_{\ell \in \{1,2,\ldots,L\}} \Omega_{1,\ell} \quad (55)$$

where $\hat{\ell} \in \{1, 2, \ldots, L\}$ is the index of the selected RT. Note that the average power of each link is assumed to be constant and does not vary with data transmissions without loss of generality. Then, the index of the selected RT will not be a discrete random variable, and the SSI-based relay selection protocol will definitely turn into the well-known AP-based partial-RSP which is also known as pathloss-based relay selection [11]. In this case, (50) reduces into the unified performance of the AP-based partial-RSP, that is

$$\mathcal{P}_{AUP} = \int_0^\infty \frac{\mathcal{Z}_{a,b}^n(u)}{\Gamma(m_{1,\hat{\ell}})\Gamma(m_{2,\hat{\ell}})} \left\{ \frac{\phi_{1,\hat{\ell}}}{\Omega_{1,\hat{\ell}}} \Gamma\left(m_{1,\hat{\ell}} - \frac{1}{\xi_{1,\hat{\ell}}}, 0, \frac{\phi_{1,\hat{\ell}} u}{\Omega_{1,\hat{\ell}}}, \frac{1}{\xi_{1,\hat{\ell}}}\right) \Gamma\left(m_{2,\hat{\ell}}, 0, \frac{\phi_{2,\hat{\ell}} u}{\Omega_{2,\hat{\ell}}}, \frac{1}{\xi_{2,\hat{\ell}}}\right) + \right.$$
$$\left. \frac{\phi_{2,\hat{\ell}}}{\Omega_{2,\hat{\ell}}} \Gamma\left(m_{1,\hat{\ell}}, 0, \frac{\phi_{1,\hat{\ell}} u}{\Omega_{1,\hat{\ell}}}, \frac{1}{\xi_{1,\hat{\ell}}}\right) \Gamma\left(m_{2,\hat{\ell}} - \frac{1}{\xi_{2,\hat{\ell}}}, 0, \frac{\phi_{2,\hat{\ell}} u}{\Omega_{2,\hat{\ell}}}, \frac{1}{\xi_{2,\hat{\ell}}}\right) \right\} du. \quad (56)$$

As a result from these analytical expressions, the performance of the SSI-based partial relay selection certainly deteriorates approaching to those of RR-based and AP-based partial-RSPs as the shadowing conditions in the fading channels from the ST to the RTs $R_1, R_2, \ldots, R_L$ get better. It is at this point worth mentioning that the shadowing conditions in all available channels from the ST to the RTs gets worse as the RTs get farther away from the ST. Then under severe shadowing conditions, the SSI-based partial relay selection turns into a feasible solution in the millimeter wave radio frequencies to maintain the performance of dual-hop AF relaying systems subjected to severe limitations such as fast the time-varying nature of the CSI and the backhaul/transmission delay in feedback channels.

*C. Numerical Results in Extended Generalized-K Fading Environments*

As an illustration of the numerical accuracy of the presented mathematical formalism, we consider the EGK fading environments for different number of RTs, i.e., $L \in \{1, 2, 3, 5\}$ with the statistical channel parameters $m_{k,\ell}$, $\xi_{k,\ell}$, $n_{k,\ell}$, $\zeta_{k,\ell}$ and $\Omega_{k,\ell}$ for all hops $k \in \{1, 2\}$ and all RTs $\ell \in \{1, 2, \ldots, L\}$ as shown in Table I. Accordingly, the ABEP performance of the SSI-based partial relay selection, obtained by simulation and numerical computations, is depicted in Fig. 2 and Fig. 3 for the BPSK and NCFSK modulations schemes, respectively, for different number of RTs. As seen in these figures, the numerical and simulation results are in perfect agreement. Further note that in our computer-based simulations, the number of samples for the instantaneous SNR $\gamma_{end}$ is chosen $10^8$ in order to obtain quite accurate results for comparison reasons.

With the fact that the performance of all RSPs significantly worsens to those of AP-based and RR-based RSPs because of some certain limitations mentioned previously, the performance of both the RR-based and AP-based partial-RSPs are also depicted in both Fig. 2 and Fig. 3. For example, the AP-based and RR-based partial-RSPs do not provide a significant performance gain while the average SNR and the number of RTs increase. On the other hand, the SSI-based partial-RSP has still good and respectable performance in shadowed fading environments. More specifically in Fig. 2 for a specific desired BPSK ABEP $2 \times 10^{-2}$ achievable with $L = 4$ number of RTs, the SNRs of the AP-based and RR-based partial RSPs are $55\,\text{dB}$ and $34\,\text{dB}$, respectively. On the other hand, the SSI-based partial-RSP reaches the same ABEP for only the average SNR $13\,\text{dB}$. In addition in order to further quantify the performance gain of the SSI-based partial-RSP and demonstrate how it turns into a feasible solution in shadowed fading environments, we also compared in both Fig. 2 and Fig. 3 its ABEP performance with that of the CSI-based partial RSP even if the CSI-based partial-RSP is somehow difficult, as discussed in more details in the introduction, to realize in millimeter wave radio frequencies. Moreover for $L = 4$ number of RTs, we can see in Fig. 3 that the NCFSK performance of the SSI-based partial-RSP is approximately $5.5\,\text{dB}$ worse than that of the CSI-based partial-RSP for the specific ABEP $10^{-3}$. However, it provides some significant strengths to some certain limitations mentioned previously. Further with respect to the estimation, synchronization as well as information combining in millimeter wave frequencies, it maintains the network complexity as simple as possible while reducing the signal processing load at the CE and increasing the overall quality of service. Note that the other information theoretical framework for studying performance limits in the SSI-based partial-RSP is the channel capacity whose first-order statistics is typically known as the EC denoting the maximum information rate for which arbitrarily small BEP can be achieved in fading environments. Having been obtained by SSI-based partial RSP, the improvement in the EC is illustrated in Fig. 4 for different number of RTs, and compared to those of the RR-based and AP-based RSPs. As seen in Fig. 4, the EC of the SSI-based partial RSP improves as the number of RTs increases.

In addition to this performance gain and complexity reduction, it is clearly expected that the performance gain of the SSI-based partial-RSP strictly depends on the random nature of the shadowing conditions. In particular, shadowing effects certainly disappear in the fading environments between the ST and RTs when the ST gets closer to the RTs. Under the same circumstances in EGK fading environment, the shadowing figures of the first-hop fading channels grow substantially (i.e., $n_{1,\ell} \to \infty$ for all $\ell \in \{1, 2, \ldots, L\}$). For 4 number of RTs under the reasonable assumption that their channels from the ST are exposed to the same shadowing distribution (i.e., $n_{1,\ell} = n$ for all $\ell \in \{1, 2, \ldots, 4\}$ and the other channel parameters are listed in Table I), Fig. 5 presents the ABEP of the SSI-based partial-RSP for the BDPSK modulation scheme with comparison to those of CSI-based, RR-based, and AP-based RSPs. Simulation and numerical curves show that the ABEP of the SSI-based partial-RSP gets gradually worsen to that of the AP-based partial-RSP as the shadowing severity in the first-hop fading channels decreases. Takin into account that performance behavior, we can quite easily deduce that the SSI-based partial-RSP certainly turns into a feasible solution in *severely shadowed* fading environments.

## V. Conclusion

In conclusion, the RSPs available in the literature are typically using the CSI. On the other hand in the millimeter wave radio frequencies, the CSI varies very fast so its coherent time is much smaller than that of the CSI in low-frequencies under the same mobility conditions. Because of some certain limitations such as backhaul/transmission delays in feedback channels, the efficient usage of CSI for RT selection may not always be possible. However and fortunately, the SSI variates more slowly, so we showed in this paper that the SSI can also be employed in the partial-RSP to obtain some respectable performance in deeply shadowed fading environments. More specifically in order to achieve the RT selection in dual-hop AF relaying system, we proposed a partial-RSP utilizing only the SSI of the fading channels from the ST to the RTs instead of the CSI of these channels. Then for the computationally efficient numerical analysis of the proposed SSI-based partial-RSP, we offered a unified and generic UP expression which simultaneously combines the ABEP analysis of binary modulation schemes (BPSK, BDPSK, BFSK and NCFSK), EC analysis, and MGF-based statistical characterization in parallel. As such, this generic UP expression can be efficiently computed for different performance analyzes and it is certainly an MGF-based approach applicable for a variety of fading environments. In addition, we reintroduced the EGK distribution for possible applications of the SSI-based RSP in composite fading environments. As an illustration of the mathematical formalism, some numerical and simulations are carried out and they are shown to be in perfect agreement.

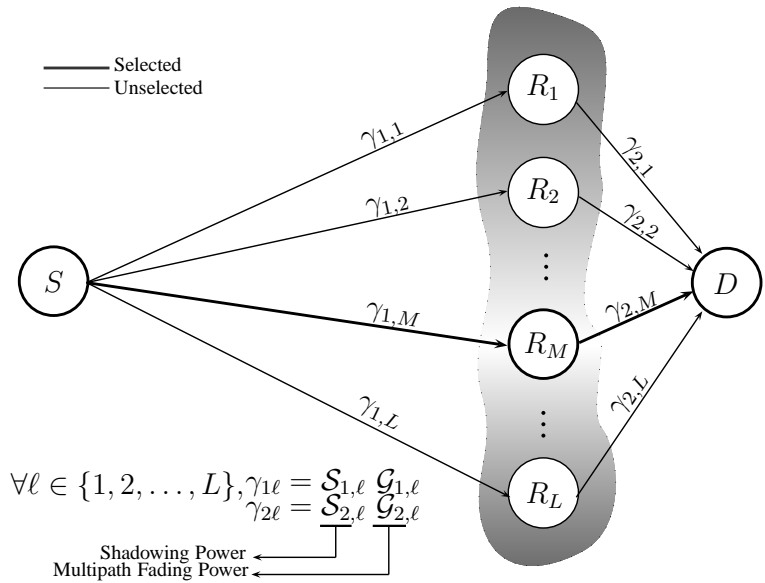

Fig. 1.   Partial relay selection with respect to shadowing side information.

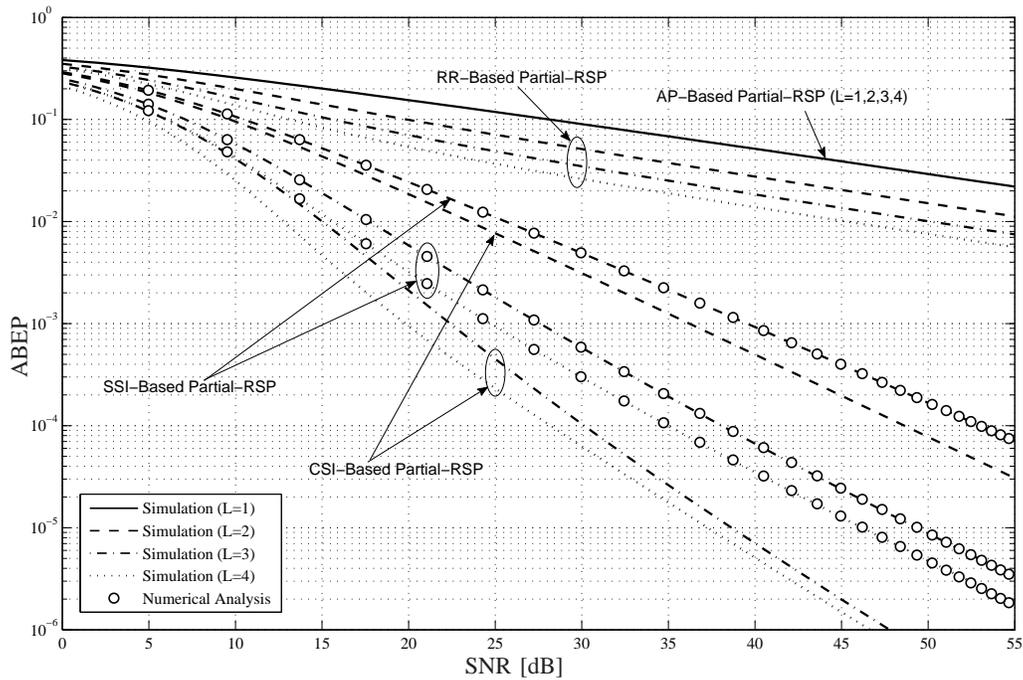

Fig. 2.   BPSK performance comparison in EGK fading environments.

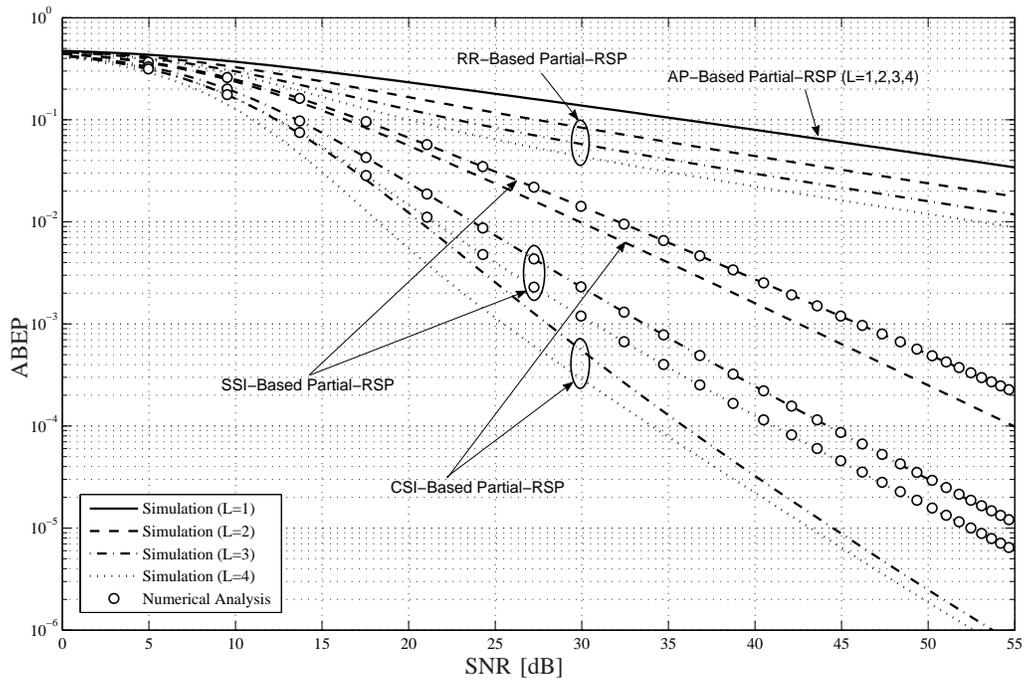

Fig. 3.  NCFSK performance comparison in EGK fading environments.

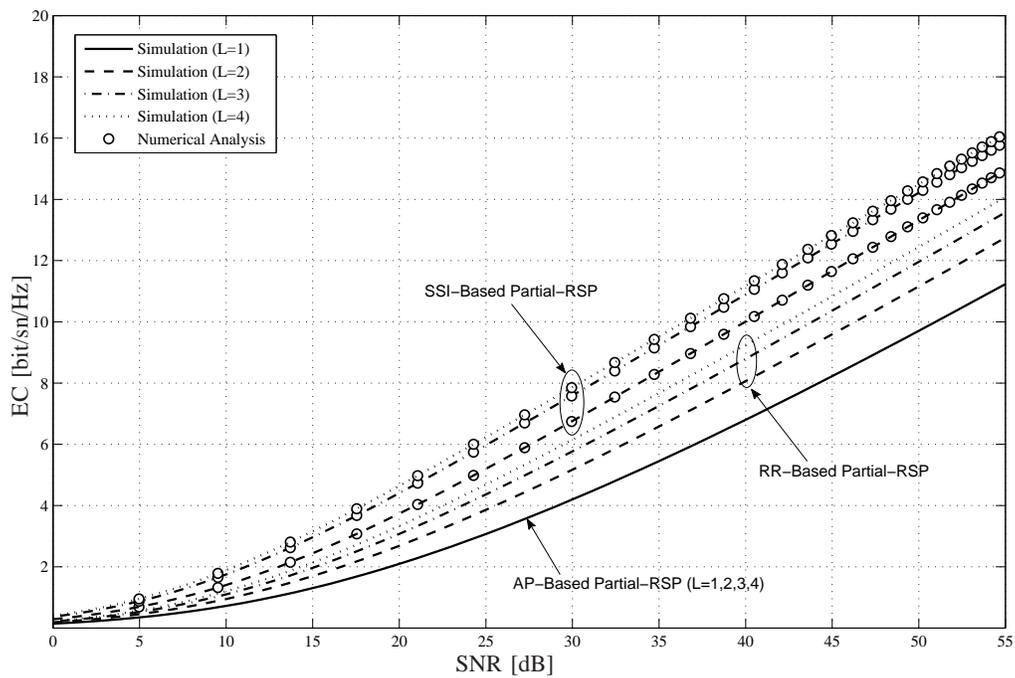

Fig. 4.  Ergodic capacity comparison in EGK fading environment.

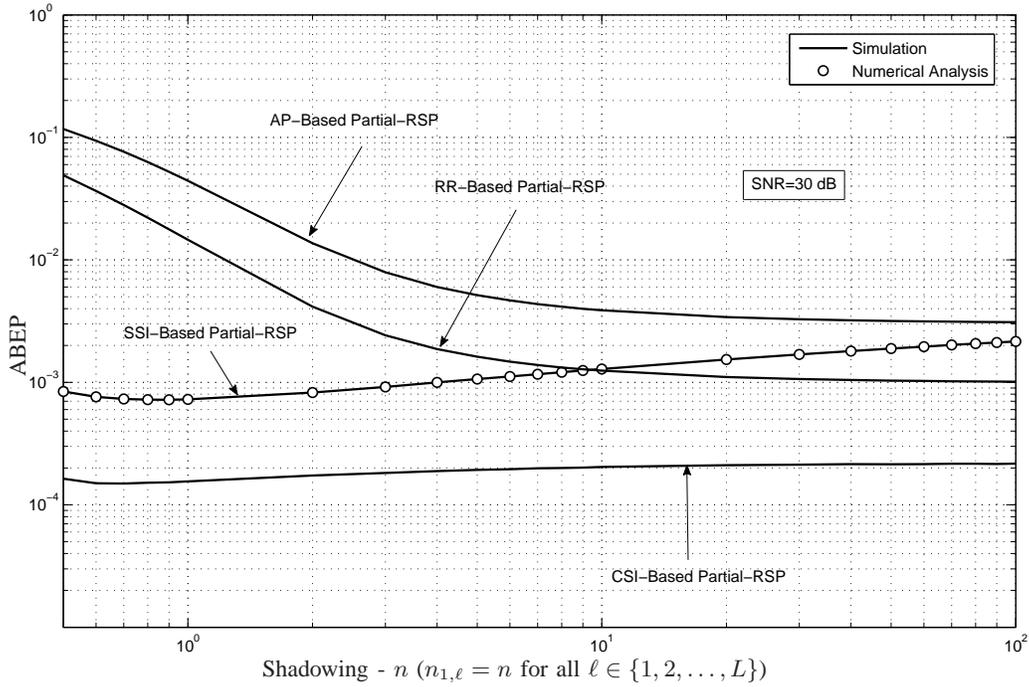

Fig. 5. For the average SNR $30$ dB and $4$ number of RTs, the BDPSK performance comparison with respect to shadowing figure of the first-hop fading channels in EGK fading environment (i.e., $n_{1,\ell} = n$ for all $\ell \in \{1, 2, \ldots, 4\}$).

TABLE I
STATISTICAL CHANNEL PARAMETERS IN DOUBLE-HOP COMMUNICATIONS OPERATING IN EGK FADING ENVIRONMENT

| Relay Terminal | Channel Hop | Fading Figure $m_{k,\ell}$ | Fading Shaping $\xi_{k,\ell}$ | Shadowing Severity $n_{k,\ell}$ | Shadowing Shaping $\zeta_{k,\ell}$ | Average Power $\Omega_{k,\ell}$ |
|---|---|---|---|---|---|---|
| 1 | 1 | 1.00 | 0.80 | 0.50 | 0.50 | 0.80 |
|   | 2 | 1.00 | 1.00 | $\infty$ | 1.00 | 0.90 |
| 2 | 1 | 1.20 | 0.90 | 0.75 | 0.75 | 0.70 |
|   | 2 | 1.25 | 1.00 | $\infty$ | 1.00 | 0.90 |
| 3 | 1 | 1.30 | 1.00 | 1.00 | 1.00 | 0.60 |
|   | 2 | 1.50 | 1.00 | $\infty$ | 1.00 | 0.90 |
| 4 | 1 | 1.40 | 1.10 | 1.25 | 1.25 | 0.50 |
|   | 2 | 1.75 | 1.00 | $\infty$ | 1.00 | 0.90 |